\pdfoutput=1

\documentclass[aps,prb,reprint,showpacs,superscriptaddress]{revtex4-1}

\usepackage{graphicx}
\usepackage{graphics}
\usepackage{amsmath}
\usepackage{amssymb}
\usepackage{amsfonts}
\usepackage{dcolumn}
\usepackage{dsfont}
\usepackage{latexsym}
\usepackage{rotating}
\usepackage{color}
\usepackage{latexsym}
\usepackage{bbm}
\usepackage{subfigure}
\usepackage{float}
\usepackage{epsfig}
\usepackage{epsf}
\usepackage{psfrag}
\usepackage{bm}
\usepackage{amsthm}
\usepackage{eucal}
\usepackage{mathrsfs}
\usepackage{url}
\usepackage{braket}

\usepackage{color} 

\usepackage{hyperref}
\hypersetup{
colorlinks=true,final=true,
        linkcolor=blue,
        citecolor=blue,
        filecolor=blue,
        urlcolor=blue,
}
\begin{document}
\title{Structural, electronic, and magnetic properties of Vanadium-based Janus dichalcogenide monolayers : A first-principles study}

\author{Dibyendu Dey}
\email{ddey3@asu.edu}
\affiliation{Department of Physics, Arizona State University, Tempe, AZ - 85287, USA}

\author{Antia S. Botana}
\email{antia.botana@asu.edu}
\affiliation{Department of Physics, Arizona State University, Tempe, AZ - 85287, USA}
\date{\today}

\begin{abstract}
The structural, electronic, and magnetic properties of VSSe, VSeTe, and VSTe monolayers in both 2H and 1T phases are investigated via first-principles calculations. The 2H phase is energetically favorable in VSSe and VSeTe, whereas the 1T phase is lower in energy in VSTe. For V-based Janus monolayers in the 2H phase, calculations of the magnetic anisotropy show an easy-plane for the magnetic moment. As such, they should not exhibit a ferromagnetic phase transition, but instead, a Berezinskii-Kosterlitz-Thouless (BKT)  transition. A classical XY model with nearest-neighbor coupling estimates critical temperatures (T$_{BKT}$) ranging from 106 K for VSSe to 46 K for VSTe. 
\end{abstract}

\maketitle
\section{Introduction} 
The discovery of long-range magnetic order in two-dimensional (2D) van der Waals (vdW) crystals has led to an upsurge in research activities on 2D magnets~\cite{GongNat, DengNat, HuangNat, HaraNL, BonillaNat}. In the past two years, single atomic layers of Fe$_3$GeTe$_2$~\cite{DengNat}, CrI$_3$~\cite{HuangNat}, MnSe$_2$~\cite{HaraNL}, and VSe$_2$~\cite{BonillaNat}, among others, have been reported to exhibit long-range magnetic order. Magnetic order in 2D can only happen if there is no continuous spin symmetry, otherwise, the proliferation of low-energy spin waves, that lies behind the Mermin-Wagner theorem~\cite{MerminPRL}, destroys magnetic order at any finite temperature. Magnetic anisotropy is hence an important requirement for realizing 2D magnetism~\cite{GongNat, DengNat}. 
vdW magnets are a perfect resort since they have an intrinsic magnetocrystalline anisotropy due to the reduced symmetry of their layered structures. 
2D materials with an easy magnetization plane should not exhibit a ferromagnetic phase transition, but instead, a Berezinskii-Kosterlitz-Thouless (BKT) transition~\cite{BKT} to a quasi-long-range ordered low-temperature phase. In contrast, 2D magnetic materials with an easy magnetization axis can exhibit a ferromagnetic (FM)  low-temperature phase. 2D vdW materials offer an additional advantage as their magnetic properties can be manipulated  via strain~\cite{MaACSN}, gating~\cite{HuangNatN, BurchNat}, or heterostructuring~\cite{GibeNat, GongSC}.

Among the above examples, the magnetic ground state of single-layer VSe$_2$  is still a matter of intense investigation and continuous debate~\cite{BonillaNat, CoelhoPCC, ChenPRL}. Recent experiments reveal the presence of a charge density wave (CDW) instability with no sign of FM ordering~\cite{CoelhoPCC, ChenPRL}, whereas earlier experimental findings claim the observation of ferromagnetism at room temperature~\cite{BonillaNat}. 2D TMDs commonly occur in two polymorphs, namely the 1T- and 2H-polytypes, in which the transition metal atoms are coordinated with the neighboring chalcogens either in an octahedral (1T) or trigonal prismatic (2H) environment~\cite{LiJPCC, Pas2D, ZhuPRB, EstPRB}. Experimentally, it has been reported that both bulk and monolayer VSe$_2$ crystallize into the 1T phase~\cite{FengNL}, with CDW getting stabilized and further enhanced in the monolayer with respect to the bulk ~\cite{CoelhoPCC, ChenPRL, FengNL}. CDW could be suppressed in the 2H phase, resulting in the stabilization of ferromagnetism in this polymorph. However, 2H-VSe$_2$ monolayer is not stable.

An open question is whether one can stabilize the 2H phase along with ferromagnetism by designing vanadium-based Janus monolayers (VXY, X/Y=S, Se, Te, and X$\neq$Y). In Janus compounds inversion symmetry is broken as different anions occupy the top and bottom layers of V atoms (see Fig.\ref{Fig1}). Prospects to grow magnetic Janus dichalcogenide monolayers are bright as non-magnetic MoSSe has already been successfully synthesized ~\cite{LuNatN}.
Some attention has been paid to V-based Janus dichalcogenide monolayers in theoretical studies~\cite{HeCMS, ZhangNL}, but VXY monolayers have not been synthesized yet. \textit{Ab initio} calculations have focused mostly on the 1T-phase~\cite{HeCMS}. Zhang {\it et al.}~\cite{ZhangNL} have analyzed the piezoelectric response and valley polarization in 2H-VSSe monolayers but have not analyzed magnetic anisotropies. VSeTe and VSTe monolayers in the 2H phase have never been discussed in the literature. These two compounds are particularly interesting as the presence of the heavy chalcogen Te should provide strong spin-orbit coupling (SOC), and enhance the magnetic anisotropy.

Here, we perform first-principles calculations to study the structural, vibrational, electronic, and magnetic properties of VSSe, VSeTe, and VSTe monolayers. An analysis of the dynamic stability of the 1T versus the 2H phase shows that the latter is dynamically stable in all cases.  2H-VXY monolayers manifest strong in-plane magnetic anisotropy and belong to the family of XY-magnets. As a consequence, they should not exhibit ferromagnetism but rather a BKT transition to a quasi-long-range ordered low-temperature phase. 

\section{Computational Details}
\label{meth}
Our density functional theory (DFT)~\cite{KohnShamPR, HohenKohnPR} calculations have been performed using a plane-wave basis set and projector-augmented wave (PAW) potentials~\cite{BlochPAW, KressePAW} as implemented in the Vienna {\it ab initio} simulation package (VASP)~\cite{KresseVASP1, KresseVASP2}. The wave functions were expanded in the plane-wave basis with a kinetic energy cutoff of 500 eV. 

 For the exchange-correlation functional, we start by using the Perdew-Burke-Ernzerhof (PBE)~\cite{GGAPBE} version of the generalized gradient approximation (GGA) to study the structural properties of Janus compounds in Section~\ref{struc_prop}. In order to better address their electronic and magnetic properties, GGA(PBE)+U using the fully localized version for the double counting correction \cite{LeichPRB} has  been used in Section~\ref{mag}. DFT+U improves over GGA or local-density approximation (LDA) in the study of systems containing correlated electrons by introducing an on-site Coulomb repulsion U~\cite{Hubbard} applied to the localized electrons (e.g. V-3$d$).
For the GGA+U calculations, we use U= 2.7 eV, reasonable for this type of 3d electron system~\cite{RPA}. A nonzero value of Hund's coupling J= 0.7 eV has been considered to account for the anisotropy of the interaction. Our choice of U and J is consistent with values used in earlier literature to study magnetism in V-based dichalcogenides within DFT+U \cite{GongPNAS,ZhuPRB,LiJPCC,ZhangNL, FuhSREP}. We note that the results presented here are consistent within the U range (2-3 eV) used in the literature. Finally, spin-orbit coupling (SOC) following its PAW implementation~\cite{Vasp_SOC} has been included (on top of GGA+U) in Section~\ref{ani} to estimate  magnetocrystalline anisotropy energies.
During structural relaxations performed within GGA, positions of the ions were relaxed until the Hellman-Feynman forces became less than 10$^{-3}$ eV/\AA. Phonons were calculated within GGA using density functional perturbation theory (DFPT) as implemented in the PHONOPY code~\cite{Phonopy}. The reciprocal space integration was carried out with a $\Gamma$-centered k-mesh of 24$\times$24$\times$1 for the conventional cell, and 12$\times$12$\times$1 for the 2$\times$2$\times$1 supercell used in phonon calculations.

\section{Results and Discussion}
\label{res}
\subsection{Structural Properties}
\label{struc_prop}
Since monolayers of vanadium-based Janus dichalcogenides have not been synthesized yet, we first benchmark the calculated in-plane lattice constant of monolayer VSe$_2$ comparing it to experiments. We use as a reference the value of single-layer VSe$_2$ grown on a bilayer graphene/SiC substrate ($a$ = 3.31 $\pm$ 0.05 \AA) for which the coupling between substrate and monolayer is weak, and hence the effect of strain is negligible \cite{FengNL}.  From FM GGA-PBE calculations, we obtain $a$ = 3.33 \AA~ for VSe$_2$ monolayers, in good agreement  with the experimentally reported value. Thus, we adopt the same procedure in calculating the in-plane lattice constants of VXY monolayers.

\begin{figure}
\includegraphics[width=8.5cm]{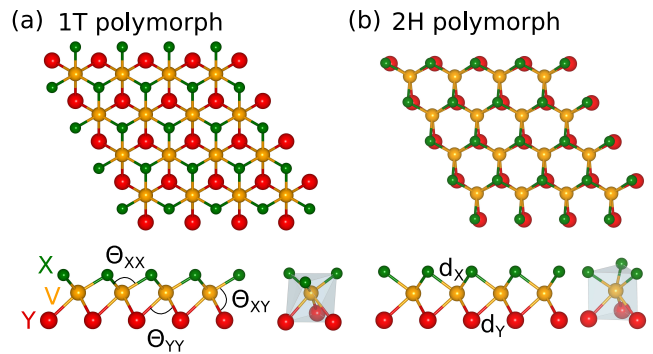}
\caption{(Color online) Crystal structure of the (a) 1T and (b) 2H polymorphs of V-based Janus (VXY) monolayers. V atoms are depicted in golden yellow, X atoms in dark green, and Y atoms in red. The X-V-X, Y-V-Y, and X-V-Y bond angles are indicated as $\theta_{XX}$, $\theta_{YY}$, and $\theta_{XY}$, respectively. The distances between V and X (Y) atoms are shown as d$_X$ (d$_Y$). The crystal field environment for V atoms in the two phases is displayed on the bottom right side of (a) and (b) panels- octahedral for the 1T phase, trigonal prismatic for the 2H phase.}
\label{Fig1}
\end{figure}

Crystal structures of VXY monolayers in both the 1T and 2H phases are depicted in Fig.~\ref{Fig1}, which shows that the atomic stacking differs in the respective polymorphs.  VXY monolayers in both  phases have P3m1 space group. In the 1T phase, V atoms are in Wyckoff position 1c (2/3, 1/3, z), X atoms in 1a (0,0,z), and Y atoms in 1b (1/3, 2/3, z).
In the 2H phase, V atoms are in Wyckoff position 1a (0,0,z), X and Y atoms in 1c (2/3, 1/3, z), with different z. In both phases, V atoms form a triangular lattice. As mentioned above, transition metal ions  exhibit octahedral and trigonal prismatic coordination in the 1T and 2H phase, respectively.  Lattice constants ($a$), bond lengths ($d_{X}$, $d_Y$), and bond angles ($\theta_{XX}$, $\theta_{YY}$, $\theta_{XY}$) are listed in Table~\ref{Tab1} after GGA-PBE relaxations in a FM state.  The trend observed in lattice constants and bond lengths agrees with the size of the anions: $r_S < r_{Se} < r_{Te}$. Among the three Janus systems, VSSe has the smallest lattice constant, whereas VSTe and VSeTe have larger values. The bond lengths follow the sequence $d_{V-S} < d_{V-Se} < d_{V-Te}$. The values of anion-V-anion bond angles are close to 90$^o$ but they are slightly different for top and bottom layers giving rise to a distorted environment for a given vanadium ion.

\begin{table*}
\begin{center}
\begin{tabular}{p{1.5cm} p{1.5cm} p{1.2cm} p{1.2cm} p{1.2cm} p{1.2cm} p{1.2cm} p{1.2cm} p{1.2cm} p{1.2cm} p{1.2cm} p{1.2cm} p{0.01cm}}
\\
\hline
 \hline
\centering & \centering Polytype & \centering $a$  & \centering $d_X$  & \centering $d_Y$  & \centering $\theta_{XX}$  & \centering $\theta_{YY}$  & \centering $\theta_{XY}$ & \centering m$_V$ & \centering m$_X$ & \centering m$_Y$& \centering $\Delta E$ &\\
\centering  & \centering & \centering (\AA) & \centering (\AA) & \centering (\AA) & \centering (deg) & \centering (deg) & \centering (deg) & \centering ($\mu_B$)& \centering ($\mu_B$)& \centering ($\mu_B$)& \centering (meV) &\\
  \hline
\centering  VSSe & \centering 1T & \centering 3.26 & \centering 2.34  & \centering 2.50 & \centering 88.2 & \centering  81.5  &  \centering 95.0 & \centering 0.70 & \centering -0.02 & \centering -0.05 & \centering 42.6 &\\
\centering  & \centering 2H & \centering 3.25 & \centering  2.35  & \centering 2.50 & \centering 87.3 & \centering  80.9  &  \centering 78.6 &
 \centering 1.00 & \centering -0.02 & \centering -0.07 & \\
\\
\centering  VSeTe & \centering 1T & \centering 3.49 & \centering 2.48  & \centering 2.72 & \centering 89.5 & \centering  79.8  &  \centering 95.1  & \centering 1.09 & \centering -0.04 & \centering -0.09 & \centering 6.1 &\\
\centering  & \centering 2H & \centering 3.46 & \centering  2.50  & \centering 2.72 & \centering 87.8 & \centering  79.1  &  \centering 79.4 & \centering 1.06 & \centering -0.03 & \centering -0.08 &\\
\\
\centering  VSTe & \centering 1T & \centering 3.46 & \centering 2.33  & \centering 2.74 & \centering 95.7 & \centering  78.1  &  \centering 92.4 & \centering 1.20 & \centering -0.03 & \centering -0.10 & \centering -30.0&\\
\centering  & \centering 2H & \centering 3.39 & \centering  2.35  & \centering 2.72 & \centering 92.2 & \centering  77.1  &  \centering 77.6 & \centering 0.79 & \centering -0.01 & \centering -0.05 &\\
  \hline
  \hline
\end{tabular}
\caption{Lattice constants ($a$), bond angles ($\theta$), bond lengths ($d$), and magnetic moments of V-based Janus monolayers after GGA relaxations in a FM state. The magnetic moments $m_V$, $m_X$, $m_Y$ are associated with V, X and Y atoms (where X/Y = S, Se, Te). $\Delta E$ denotes the energy difference between the 1T and 2H phases per formula unit (f.u.) -negative energies obtained when the 1T phase is more stable.}
\label{Tab1}
\end{center}
\end{table*}
\begin{figure}
\begin{center}
\includegraphics[width=8.2cm]{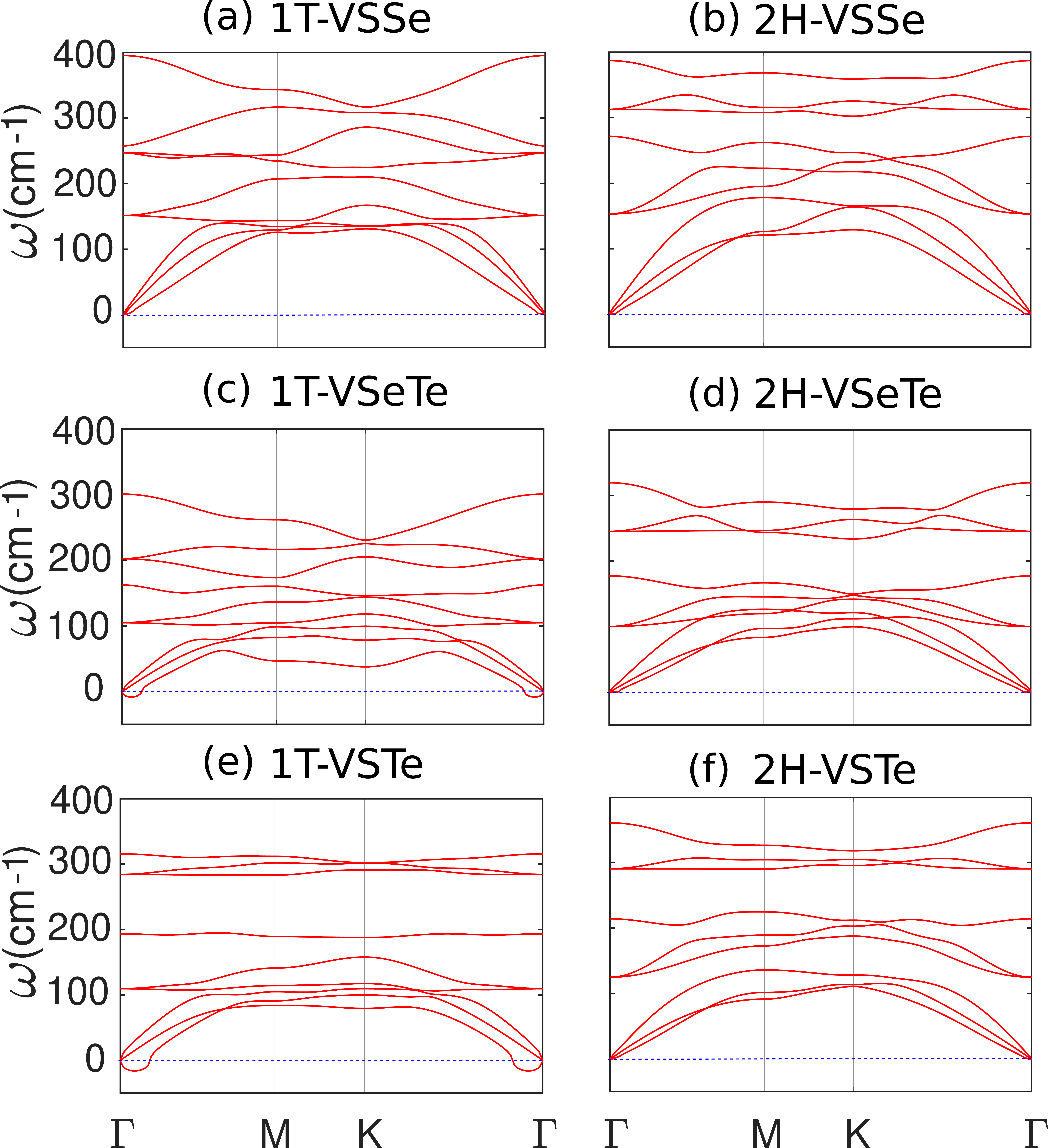}
\caption{\label{Fig2}(Color online) Phonon dispersions from GGA calculations in a FM state for VSSe (a,b), VSeTe (c,d), and VSTe (e,f) monolayers in 1T (left panel) and 2H (right panel) phases along the high-symmetry direction $\Gamma-$M$-$K$-\Gamma$.}
\end{center}
\end{figure}

In order to address the dynamic stability of VXY monolayers in both the 1T and 2H phases, we have performed phonon calculations within GGA-PBE in a FM state. The 2H phase is dynamically stable as no imaginary frequency has been observed in the phonon dispersions (see Fig.~\ref{Fig2} (b), (d), (f)). Monolayer 1T-VSSe does not show any imaginary modes in the phonon dispersion either (Fig.~\ref{Fig2}(a)). For 1T-VSeTe and 1T-VSTe, calculations reveal that one of the acoustic phonon branches along the $\Gamma$-M and $\Gamma$-K directions has imaginary frequencies, although real frequencies are found for all the optical modes (see Fig.~\ref{Fig2} (c), (e)). The instability close to the $\Gamma$ point persists regardless of the exchange-correlation functional used, size of the supercell, k-mesh, and size of the density grid.   

Further, we have calculated the energy difference $\Delta E = E_{1T}-E_{2H}$ within GGA-PBE between the 1T and 2H phases in a FM state to find the energetically stable polymorph of VXY monolayers. The energy differences are listed in Table~\ref{Tab1}. Details on the corresponding GGA electronic structures are provided in the next section. For VSSe and VSeTe monolayers, the 2H phase is lower in energy than the 1T phase with a $\Delta E$ of 42.6 meV/f.u. and 6.1 meV/f.u., respectively. In a recent study, Zhang {\it et al.}~\cite{ZhangNL}  also found the 2H polytype of VSSe is energetically more favorable. On the contrary, for VSTe monolayers, the 1T phase is found to be more stable with a $\Delta E$ of -30.0 meV/f.u. in spite of the above-mentioned imaginary modes appearing close to $\Gamma$.  In recent studies, a CDW phase has been confirmed in 1T-VSe$_2$~\cite{CoelhoPCC, ChenPRL, FengNL} and 1T-TiSe$_2$~\cite{SugaACSNano,SinghPRB} monolayers. Based on this, the results we find for 1T-VSTe monolayers suggest that there might be spontaneous relaxation into a CDW phase in this system as well. Experiments would be required to confirm this possibility in 1T-VSTe and to provide guidance on the periodicity of the CDW.

\begin{figure}
\begin{center}
\includegraphics[width=0.9\columnwidth]{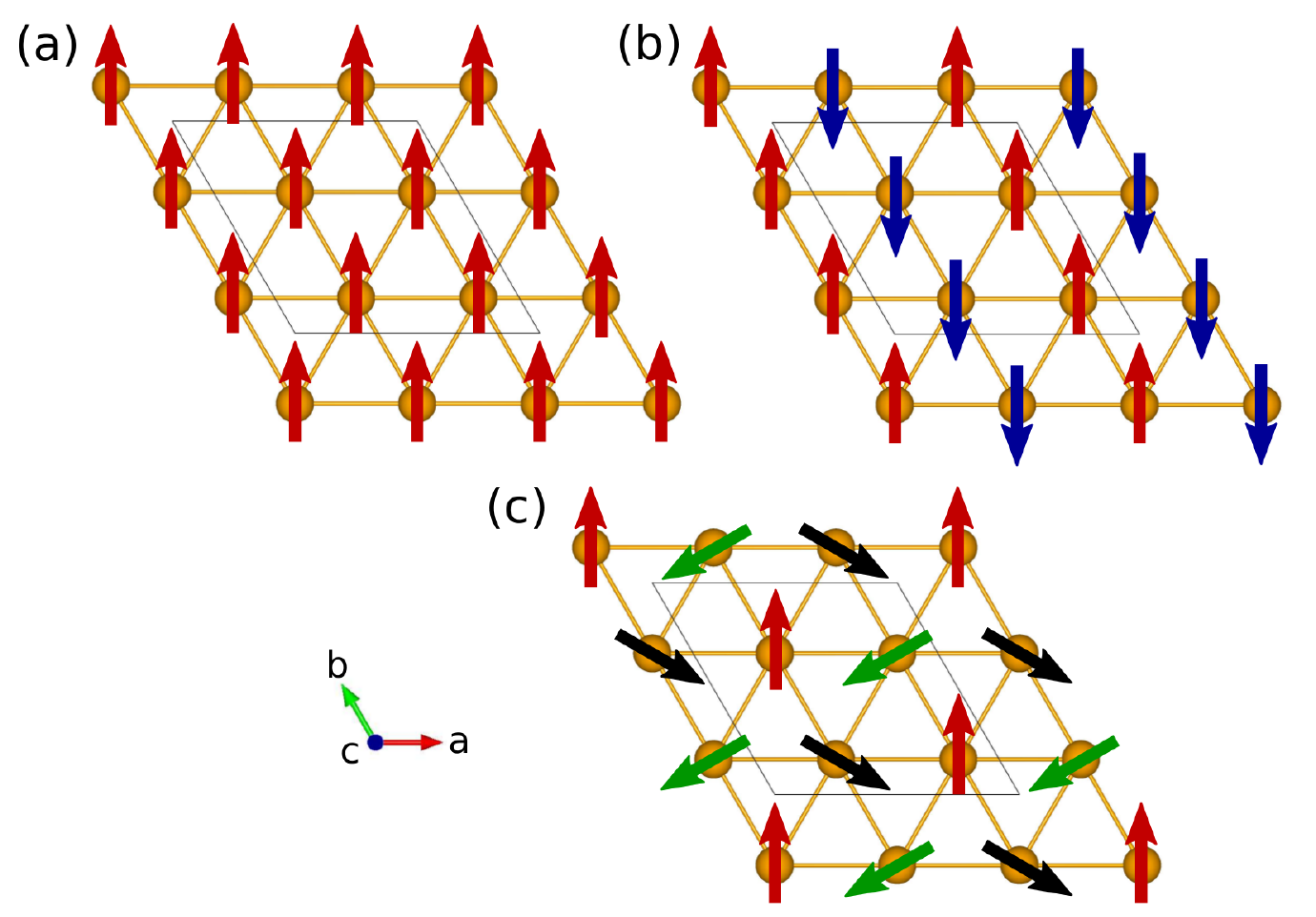}
\caption{\label{Fig3}(Color online) Top view of the effective triangular lattice formed by V ions in VXY monolayers showing (a) a ferromagnetic state, (b) a stripe antiferromagnetic state (AFM$_1$), and (c) a non-collinear 120$^o$ antiferromagnetic state (AFM$_2$). V ions are shown as golden spheres.}
\end{center}
\end{figure}

\subsection{Electronic and Magnetic Properties}
\label{mag}

\begin{figure}
\begin{center}
\includegraphics[width=8.1cm]{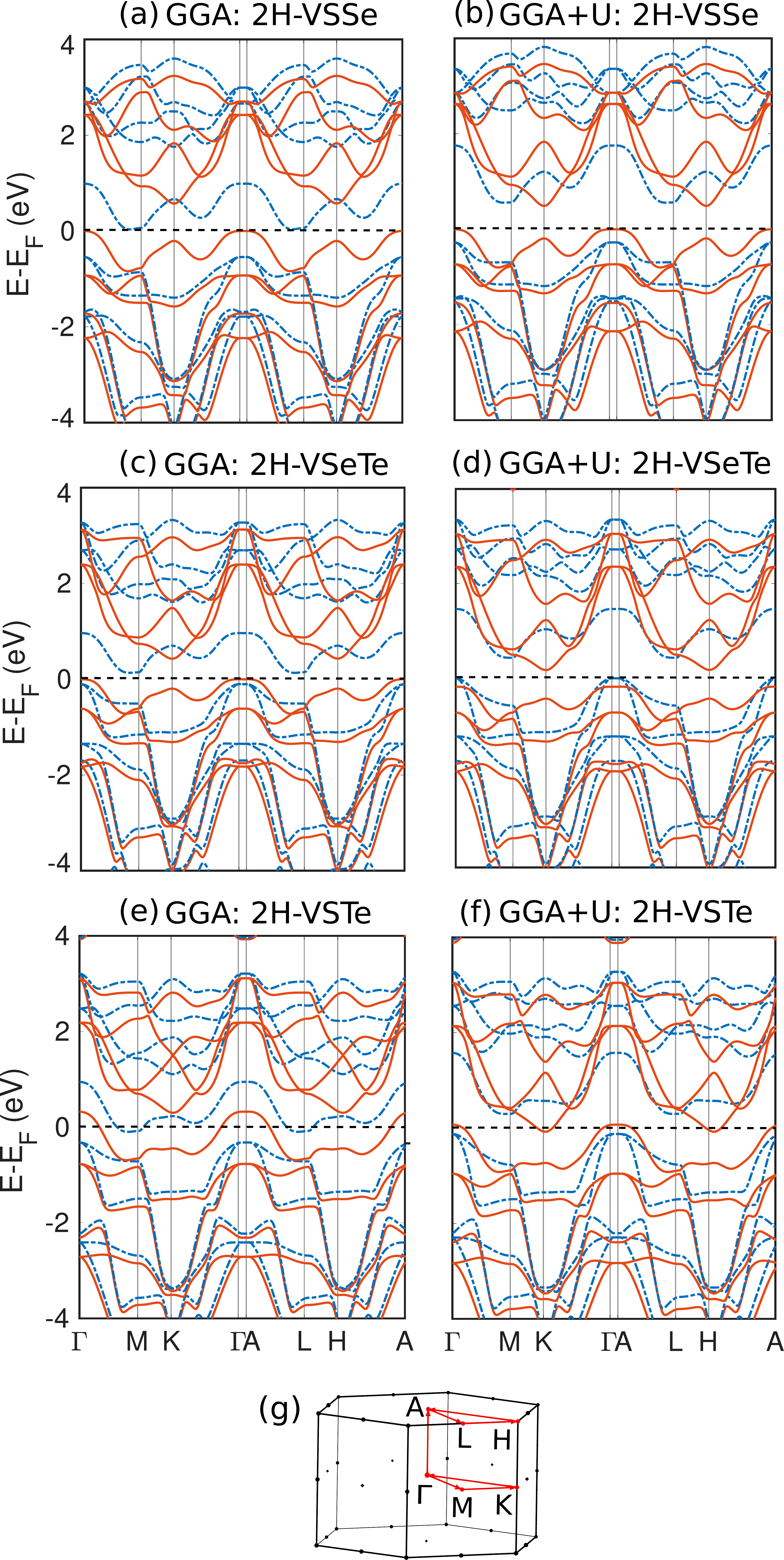}
\caption{\label{Fig4}(Color online) Electronic band structures in the FM state for 2H-VSSe (a,b), 2H-VSeTe (c,d), and 2H-VSTe (e,f) monolayers within GGA (left panels) and GGA+U (right panels). Red solid lines represent the majority spin channel, and blue dotted lines  the minority spin channel. (g) Brillouin zone of 2H-VXY monolayers showing the high-symmetry points used in the bandstructure plots: $\Gamma$=(0,0,0), M=(1/2,0,0), K=(1/3,1/3,0), A=(0,0,1/2), L=(1/2,0,1/2), and H=(1/3,1/3,1/2).}
\end{center}
\end{figure}

 In order to account for on-site correlations and to be able to compare the FM configuration with other magnetic states, we have performed GGA+U calculations for 2H-VSSe, VSeTe, and VSTe monolayers as well as for single-layer 1T-VSTe. In addition to the FM state, two standard antiferromagnetic (AFM) spin configurations for a triangular lattice have been used: a stripe AFM state (AFM$_1$), and non-collinear 120$^o$ AFM order (AFM$_2$). The present theoretical consensus is that the latter is the ground state for a spin-1/2 triangular lattice Heisenberg antiferromagnet~\cite{Heisen1, Heisen2}. A schematic representation of the three magnetic states used is depicted in Fig.~\ref{Fig3} (a)-(c). 
 We note that the introduction of an on-site U is necessary to stabilize an AFM state: without a U, vanishing moments are obtained.

\begin{table}
\begin{center}
\begin{tabular}{p{1.0cm} p{1.0cm} p{1.0cm} p{1.0cm} p{1.0cm} p{1.0cm} p{1.5cm} p{0.01cm}}
\\
\hline
\hline
\centering  & \centering & \centering & \centering m$_V$ ($\mu_B$)& \centering m$_X$ ($\mu_B$)& \centering m$_Y$ ($\mu_B$)& \centering $\Delta E_{m}$ (meV) &\\
  \hline
\centering  & \centering  & \centering FM & \centering 1.18 & \centering -0.13 & \centering  -0.07 & \centering 0.0 &\\
\centering VSSe & \centering 2H & \centering AFM$_{1}$ &
 \centering 0.48 & \centering -0.01 & \centering -0.01 & \centering 82.5 &\\
 \centering  & \centering & \centering AFM$_2$ &
 \centering 0.64 & \centering -0.00 & \centering -0.00 & \centering 190.6 &\\
\hline
\centering  & \centering  & \centering FM & \centering 1.28 & \centering -0.10 & \centering -0.16 & \centering 0.0 &\\
\centering VSeTe & \centering 2H & \centering AFM$_{1}$ &
 \centering 0.88 & \centering -0.03 & \centering -0.03 & \centering 63.5 &\\
 \centering  & \centering & \centering AFM$_2$ &
 \centering 1.42 & \centering -0.00 & \centering -0.00 & \centering 151.5 & \\
 \hline 
 \centering  & \centering  & \centering FM & \centering 1.19 & \centering -0.06 & \centering -0.15 & \centering 0.0 &\\
\centering VSTe & \centering 2H & \centering AFM$_{1}$ &
 \centering 0.75 & \centering -0.02 & \centering -0.03 & \centering 36.4 &\\
 \centering  & \centering & \centering AFM$_2$ &
 \centering 1.26 & \centering -0.00 & \centering -0.00 & \centering 117.1 &\\
  \hline   
\centering  & \centering  & \centering FM & \centering 1.67 & \centering -0.06 & \centering -0.15 & \centering 16.8 &\\
\centering VSTe & \centering 1T & \centering AFM$_{1}$ &
 \centering 1.59 & \centering -0.02 & \centering -0.04 & \centering 4.2 &\\
 \centering  & \centering & \centering AFM$_2$ &
 \centering 1.58 & \centering -0.00 & \centering -0.00 & \centering 0.0 &\\
  \hline   
  \hline
\end{tabular}
\caption{Magnetic moments and energy differences per formula unit ($\Delta E_m$) within GGA+U for the three different spin configurations  depicted in Fig. 3: FM, AFM$_1$ (stripe phase), and AFM$_2$ (non-collinear 120$^o$ spin ordering). A zero value of $\Delta E_m$ represents the magnetic ground state.}
\label{Tab2}
\end{center}
\end{table}

Magnetic moments and energy differences ($\Delta E_m$) with respect to the corresponding magnetic ground state within GGA+U are listed in Table~\ref{Tab2}. In the 2H phase, a FM configuration is always the ground state of the system. In the 1T phase of VSTe, a 120$^\circ$ non-collinear configuration is the most stable one instead. Dichalcogenide VXY monolayers are covalent in nature, and the magnetic moments (m$_V$) of the V atoms in these compounds can vary strongly. Nevertheless, m$_V$ values agree qualitatively with the ionic description that gives a V$^{4+}$: 3$d^1$ electronic configuration. For the FM state, a clear increase in the magnetic moments is obtained within GGA+U with respect to the GGA values, as expected (see Tables \ref{Tab1} and \ref{Tab2}).

As described above, VSSe and VSeTe monolayers are energetically and dynamically stable in the 2H phase, and this is also a dynamically stable polymorph for VSTe. As in the 2H phase, a  FM  configuration  is  always  the  ground  state  of  the system within both GGA and GGA+U, we compare the GGA and GGA+U band structures for all 2H-VXY monolayers to draw a consistent picture (see Fig~\ref{Fig4}). The GGA band structures of VSSe and VSeTe show a small energy gap E$_g$ = 0.04 eV (Fig.~\ref{Fig4}(a)) and E$_g$ = 0.13 eV (Fig.~\ref{Fig4}(c)), respectively. On the contrary, single-layer VSTe is metallic (Fig.~\ref{Fig4}(e)). Within GGA+U, the band gap increases in VSSe (Fig.~\ref{Fig4}(b)) and VSeTe (Fig.~\ref{Fig4}(d)) monolayers to values of E$_g$ = 0.51 eV and 0.19 eV, respectively. VSTe remains metallic (Fig.~\ref{Fig4}(f)) even within GGA+U with electron and hole pockets centered at the K (H) and $\Gamma$ (A) points, respectively. 

\begin{figure}
\begin{center}
\includegraphics[width=8.7cm]{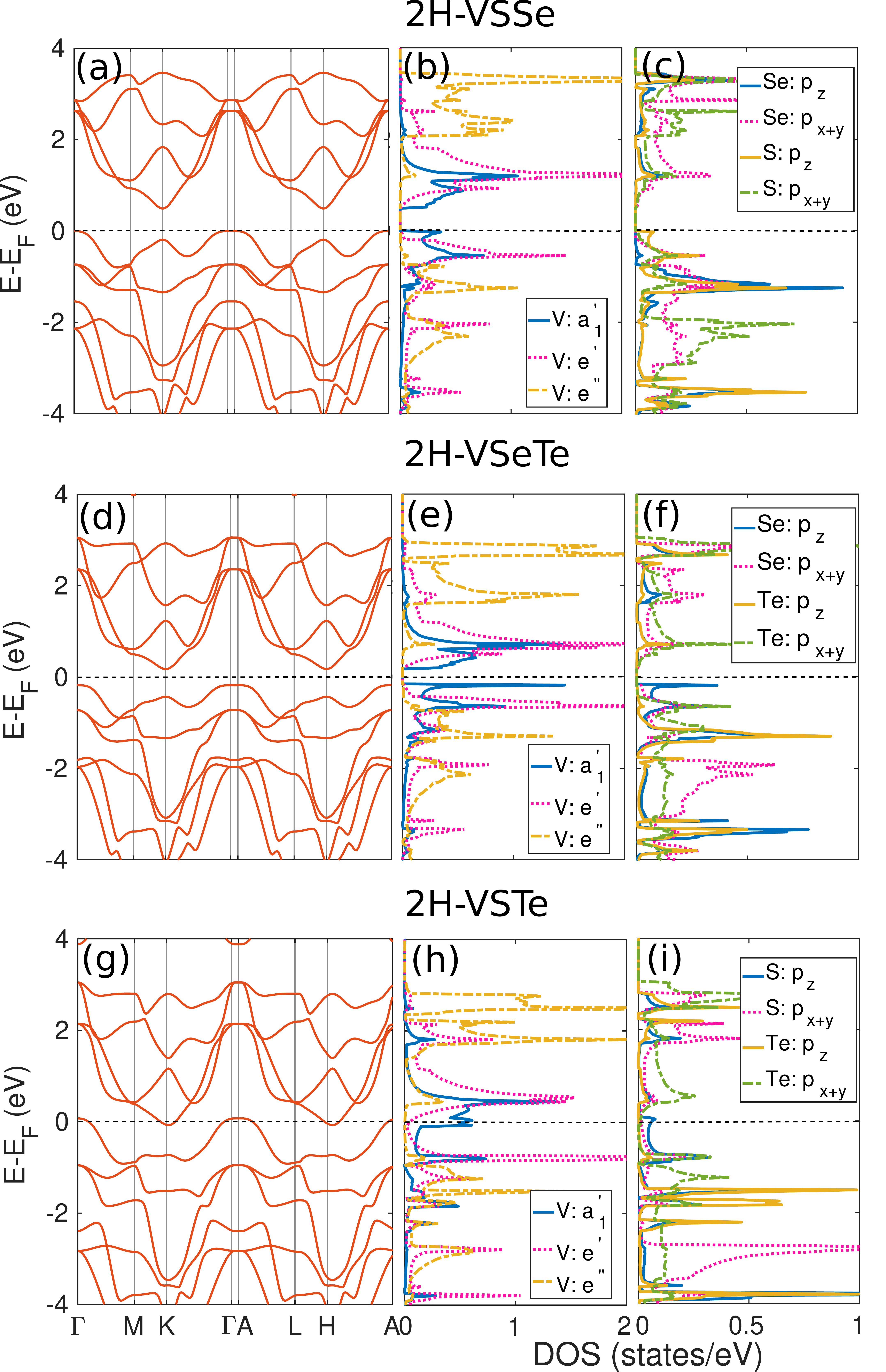}
\caption{\label{Fig5} Majority-spin band structures (right panels) and orbital projected DOS for V d-states (middle panels) and anion (S, Se, Te) p-states (right panels) from FM GGA+U calculations in 2H-VXY monolayers. (a)-(c) correspond to plots for 2H-VSSe, (d)-(f) for 2H-VSeTe, and (g)-(i) for 2H-VSTe.}
\end{center}
\end{figure}

 Fig. \ref{Fig5} shows the majority spin channel of the GGA+U band structure and the corresponding orbital-resolved V-d and chalcogen-p density of states (DOS) for 2H-VXY monolayers in a FM state. In the 2H-phase, the trigonal prismatic environment of the V atom allows splitting of the d orbitals into a lower-lying a$_1'$ ($d_{z^2}$) singlet followed by a doubly degenerate e$'$ orbital ($d_{xy}+d_{x^2-y^2}$), and a high-energy doublet e$''$ ($d_{xz}+d_{yz}$)~\cite{LiWFPRB}. In a simple ionic picture for a d$^1$ ion, one would expect a single electron in the lower-lying a$_1'$ orbital for the majority spin channel. However, this is a covalent compound, and the V-d DOS (Fig.~\ref{Fig5} (b), (e), (h)) reveals a high degree of admixture of different d-orbitals as well as chalcogen p-hybridization- even though a$_1'$-character is still dominant around the Fermi level. Mattheiss first reported this fact in the context of band structure calculations of TMDs~\cite{MattheissPRB}. The band gap opens up in VSeTe and VSSe between states that are predominantly $d$ character on both sides of the gap. It is also evident from the DOS that the e$'$ and e$''$ orbitals hybridize strongly with the p$_x$ and p$_y$ orbitals to form bonding and anti-bonding states, while the a$_1'$ orbital hybridizes with the p$_z$ orbitals.


We now turn to the nature of the obtained magnetic ground states in 2H- and 1T-VXY monolayers, which is in agreement with the Goodenough-Kanamori (GK) rules~\cite{Good, Kana}. The anion-V-anion bond angles are close to 90$^o$ as described in Table~\ref{Tab1}).  In this situation, superexchange is always FM \cite{Good, Kana}. However, depending on the active d-orbitals, direct-exchange may become important and compete with FM superexchange. For e$_g$ orbitals, with the lobes of the wavefunctions directed toward the oxygens, the hopping via oxygens is still more important, and FM wins~\cite{khom_book}. This is what happens in the 2H phase that has contributions around the Fermi level mainly from the e$_g$-like (a$_1'$) orbital, as described above. However, the situation is different if t$_{2g}$ orbitals participate in the exchange. These orbitals on neighboring sites are directed towards each other (as they point between the oxygens), giving rise to AFM direct exchange, which can overcome FM superexchange~\cite{khom_book, YarePRB}. This is what happens in the 1T-phase of VSTe for which an AFM ground state is obtained.

\subsection{Magnetic anisotropy}
\label{ani}
In 2D materials, magnetic anisotropy is an essential prerequisite for stabilizing FM, as described in the introduction. Magnetic anisotropy energies are mainly governed by the magnetocrystalline anisotropy derived from spin-orbit coupling (SOC). In order to determine the easy-axis or easy-plane nature of the spin anisotropy in FM 2H-VXY monolayers, we perform GGA+U+SOC calculations and calculate the magnetocrystalline anisotropy energies, i.e. how the direction of the spin of the V atom affects the energy when spin-orbit coupling interactions are considered. The obtained $\Delta E_{a}$ ($E_{in-plane}$-$E_z$) is -0.6 meV/f.u. for 2H-VSSe, -1.89 meV/f.u. for 2H-VSTe, and -2.34 meV/f.u. for 2H-VSeTe monolayers. All three systems have an easy magnetization plane as the rotation of the magnetic moment within the plane of the 2D layer requires no energy. Thus, vanadium-based Janus dichalcogenides monolayers in the 2H phase manifest an easy magnetization plane for spins, and they belong to the class of XY-magnets. The in-plane anisotropy is higher in 2H-VSTe and 2H-VSeTe due to the presence of the heavy chalcogen Te.  The band structures within GGA+U+SOC differ from the GGA+U ones in Fig.\ref{Fig4} only in some degeneracy liftings at $\Gamma$ with the magnetic moments and band gap values (for VSSe and VSeTe) remain identical.

XY magnets show quasi long-range order at low temperatures~\cite{BKT}. The transition from the high-temperature disordered phase to this low-temperature quasi-ordered state is known as the BKT transition~\cite{BKT, BKT2}. The corresponding transition temperatures ($T_{BKT}$) can be calculated by using the formula $T_{BKT} =0.89J/k_\beta$~\cite{FernPRB}, where J is the magnetic exchange term between neighboring spins and k$_\beta$ the Boltzmann constant. In a FM configuration, each V has six neighbors with the same spin (Fig~\ref{Fig3} (a)), and the magnetic energy for this configuration is $E_{FM}=-6J$. On the other hand, the AFM$_1$ configuration gives $E_{AFM}=2J$ as four neighbors have opposite spins, and two are having the same spin (Fig~\ref{Fig3} (b)). Hence, the exchange term can be estimated from the energy difference $\Delta E_{m} = 8J$~\cite{ZhuPRB} between AFM$_1$ and FM spin configurations listed in Table~\ref{Tab2}. We obtain a nearest-neighbor FM exchange interaction J = 10.3 meV, 7.9 meV, 4.5 meV for 2H-VSSe, 2H-VSeTe, and 2H-VSTe monolayers, and corresponding transition temperatures $T_{BKT}$ of 106 K, 82 K, 46 K, respectively. 

\section{Conclusions}
\label{conc}
 We have studied the structural, vibrational, electronic, and magnetic properties of vanadium-based Janus dichalcogenides at the monolayer level using density-functional theory-based calculations. After investigating two possible polymorphs (octahedral 1T and trigonal prismatic 2H) at the monolayer level, we have shown that the 2H phase is energetically more favorable in VSSe and VSeTe, whereas the 1T phase is lower in energy for single-layer VSTe. The 2H phase is dynamically stable in all three monolayers, but an instability appears close to the $\Gamma$ point in 1T-VSTe. This instability might indicate spontaneous relaxation into a CDW phase, a point that will have to be confirmed by future experiments. A semiconducting FM ground state is obtained in 2H-VSSe and VSeTe monolayers, whereas 2H-VSTe shows a ferromagnetic metallic ground state. Inclusion of spin-orbit coupling reveals an easy magnetization plane for spins in 2H-VXY monolayers, putting them in the class of XY-magnets. Our findings suggest that a BKT transition could occur in the 2H phase, and a classical XY model with nearest-neighbor coupling estimates critical temperatures, T$_{BKT}$ of 106 K, 82 K, and 46 K for 2H-VSSe, VSeTe, and VSTe monolayers, respectively. Our results provide guidance for new exploration, both experimental and theoretical, on vanadium-based Janus dichalcogenide monolayers.

\section{Acknowledgments}
ASB acknowledges NSF-DMR grant 1904716. DD acknowledges ASU for startup funds. We acknowledge the ASU Research Computing Center for HPC resources.

%

\end{document}